# Tunable Band gap of Iron-Doped Lanthanum-Modified Bismuth Titanate Synthesized by the Thermal Decomposition of a Secondary Phase

**Jun Young Han and Chung Wung Bark**

*Department of Electrical Engineering, Gachon University, Seongnam, Gyeonggi-do 461-701, Republic of Korea*

The photoelectric properties of complex oxides have prompted interest in materials with a tunable band gap, because the absorption  The substitution of iron atoms in La-modified bismuth titanate (BLT) can lead to dramatic improvements in the band gap, however, the substitution of iron atoms in BLT without forming a $BiFeO_3$ secondary phase is quite challenging. Therefore, a series of Fe-doped BLT (Fe-BLT) samples were characterized using a solid reaction at various calcination temperatures (300~900°C) to remove the secondary phase. The structural and optical properties were analyzed by X-ray diffraction and ultraviolet-visible absorption spectroscopy. This paper reports a new route to synthesize a pure Fe-BLT phase with a reduced optical band gap by high temperature calcination due to the thermal decomposition of $BiFeO_3$ during high temperature calcination. This simple route to reduce the second phase can be adapted to other complex oxides for use in emerging oxide optoelectronic devices.

PACS numbers: 81.05.Je, 78.20.Jq, 77.80.-e

Keywords: Bandgap, Bismuth titanate, Calcination temperature, Thermal decomposition, $BiFeO_3$

# I. INTRODUCTION

The photovoltaic effect of ferroelectric perovskite oxides has been observed in complex oxide materials [1-2]. For several decades, the photovoltaic effect observed in ferroelectric perovskite thin films has attracted significant attention because of device miniaturization and large polarization as well as its possible coupling with the magnetic moment [3-4].

Studies of the photoelectric properties of complex oxides have focused on materials with a narrow band gap because the absorption efficiency of a wide spectrum of solar radiation is limited by the band gap of the materials. For example, $BiFeO_3$ has attracted considerable interest because of its relatively low band gap of 2.7 eV compared to other perovskite oxides, such as $BaTiO_3$ (3.2eV), $LiNbO_3$ (4 eV) or $Pb(Zr,Ti)O_3$ (3.6 eV) [5-6]. On the other hand, the band gaps of these materials are much larger than that of commercially available silicon (1.1 eV).

The site-specific substitution, i.e. doping, are used widely to enhance the optical properties in complex oxides materials. For example, the correlation between the optical properties of BLT films with a variation in the La concentration were reported [7]. The refractive index decreased with increasing La concentration over the wavelength range, 250-800 nm. The refractive index can affect the crystallinity and electronic band structure or lattice point defects [8]. Substitution also can alter the band gap and Fermi level, and affect the optical response. On the other hand, La substitution in BLT films has a huge influence on the ferroelectric and optical properties but the impact on the band structure is very small. As a result, La-modified BLT has a wide band gap of more than 3.0 eV [9].

To solve the wide band gap problem in complex oxides, the site-specific substitution of metal atoms in anisotropic oxides, which can provide an opportunity to tune the band gap in transition metal oxides (TMOs) were introduced recently [10]. Substitution in complex oxides can induce a decrease in the band gap due to compositional changes. Substitution, i.e. doping, can also create localized states in the band gap, which will lead to a shift in the absorption edge towards a lower photon energy, and a decrease in the optical energy gap [11]. Layered perovskite oxides are some of the best candidate materials among a range of various complex oxides for reducing the wide band gap because of their exceptional stability with site-specific substitution. Recently, W. S. Choi et al. reported that superlattice films alloying $LaTMO_3$ (TM = Ti, V, Cr, Mn, Co, Ni, Al) with $Bi_4Ti_3O_{12}$ (BiT) can be one approach to systematically lower the band gap of a ferroelectric [12].

Information about the optical band gap reduction of BLT ($Bi_{3.25}La_{0.75}Ti_3O_{12}$) in simple solid state synthesis is incomplete. In previous studies, an attempt was made to reduce the optical band gap through cobalt or iron doping on lanthanum-modified $Bi_4Ti_3O_{12}$-based oxides (BLT) synthesized using a simple chemical reaction method [13-14]. The results showed that Co or Fe-doping of BLT could decrease the band gap dramatically without breaking their crystallographic symmetry. On the other hand, X-ray diffraction (XRD) of the Fe doped BLT samples showed that a huge amount of secondary phase ($BiFeO_3$) had formed from Bi and Fe, whereas no secondary phases formed in cobalt-doped BLT.

The substitution of iron atoms in BLT without forming a secondary phase is very challenging, because the BLT and $BiFeO_3$ begin to crystallize at similar temperatures during calcination. On the other hand, $BiFeO_3$ phase tends to decompose at high temperatures, whereas the BLT phase is still stable at the same temperature. In this study, high temperature calcination was used to find the appropriate conditions for obtaining pure Fe-BLT phases with a reduced optical band gap optimization in the phase. All samples were synthesized by a solid-state reaction followed by heat treatment at different temperatures (300~900°C). The synthesized samples were characterized structurally by XRD and scanning electron microscopy (SEM). The optical band gap of the synthesized powers was estimated by ultraviolet-visible (UV-Vis) absorption spectroscopy.

A pure Fe-BLT phase with a reduced band gap was synthesized by the decomposition of the secondary phase ($BiFeO_3$) at high temperatures. Fe-BLT calcined at 900°C showed an orthorhombic structure without a second phase and its optical band gap (~2.0 eV) was much lower than the value previously reported for BLT (~2.8 eV).

## II. EXPERIMENTAL PROCEDURE

To analyze the effects of the calcination temperature on the structure and optical band gap corresponding to the compositions, the $Bi_{3.25}La_{0.75}FeTi_2O_{12}$ (Fe-BLT) samples were calcined at temperatures ranging from 300 to 900°C. Figure 1 summarizes the entire procedure.

The samples were synthesized by a solid-state reaction using the following starting binary oxide powders: $Bi_2O_3$ (99.9%, Kojundo), $TiO_2$ (99.99%, Kojundo), $La_2O_3$ (99.99%, Kojundo), and $Fe_2O_3$ (99.9%, Kojundo). The powders were blended thoroughly at the stoichiometric ratio in a ball mill for 24 hours, dried in an oven at 100°C and calcined at different temperatures for 2.5 hours in air.

The samples were characterized structurally by XRD (Rigaku, D/MAX 2200) using Cu Kα radiation from 20 to 60° 2θ with an angular step of 0.02°/min. The microstructure of the synthesized powers was examined by SEM (Hitachi, S-4700). The optical properties were determined by UV-Vis spectroscopy (Agilent, 8453).

## III. RESULT AND DISCUSSION

Figure 2 shows XRD patterns of the Fe-BLT powders calcined at temperatures between 300 and 900°C. All samples were identified using the data from the JCPDS Powder Diffraction File and Inorganic Crystal Structure Database.

The XRD peaks of Fe-BLT powders calcined at 300, 400°C and 500°C were indexed to bismuth oxide (α-$Bi_2O_3$) and iron oxide (α-$Fe_2O_3$) with a monoclinic structure, which is in agreement with the respective JCPDS Cards No.71-0465 and No 85-0987 [15-16]. This might be the result of α-$Bi_2O_3$ phase that is stable at low-temperature and α-$Fe_2O_3$, which also has chemical stability and durability [15,17]. With further increases in the calcination temperature (from 500 to 600°C), the orthorhombic structure of BLT began to form from the reduction of α-$Bi_2O_3$, α-$Fe_2O$ phases, and appeared on the XRD peak of the BLT phase but it was still a mixture.

The XRD patterns of the powders calcined at 700, 800°C and 900°C matched those reported for $Bi_{3.25}La_{0.75}Ti_3O_{12}$ with an orthorhombic structure (ICSD No. 150091). The $BiFeO_3$ phase was strongly observed at 700 and 800°C, which was identified from JCPDS card no. 71-2494 [18]. This was attributed to the formation of a $BiFeO_3$ phase from Bi and Fe. The XRD peak intensity of $BiFeO_3$ began to decrease from 700 to 800°C with an increase in the intensity of the (117) peak of Fe-BLT phase. Pure Fe-BLT phase formation without a $BiFeO_3$ secondary phase was observed at 900°C. According to the literature [19-20], $BiFeO_3$ undergoes a phase transition from a ferroelectric α-$BiFeO_3$ phase to a paraelectric β-$BiFeO_3$ phase from 820 to 830°C. Subsequently, $BiFeO_3$ begins to decompose to $Fe_2O_3$ and liquid $Bi_2O_3$ (830~960°C). Consequently, the formation of pure Fe-BLT phase is believed to be caused by the decomposition of $BiFeO_3$ from 800 to 900°C.

The morphological and microstructural properties of the samples were examined by SEM. Figure 3 shows SEM images of the samples calcined at different temperatures between 300 to 900 °C. All the specimens consisted of plate-like grains with a random orientation of plate faces because of their highly anisotropic crystal structure. Plate-like grain formation was reported to be a typical characteristic of bismuth layer-structured ferroelectrics. The microstructure showed a round stick shape at low calcination temperatures (300, 400, 500°C), whereas the samples calcined at 600 to 900°C showed rod shapes that decreased in size with increasing temperature. This shape would be formed by $Fe_2O_3$ because in the recent synthesis of BLT ($Bi_{3.25}La_{0.75}Ti_3O_{12}$) and cobalt-doped BLT ($Bi_{3.25}La_{0.75}CoTi_2O_{12}$), stick-shaped particles were not observed in these samples by SEM [13-14]. XRD showed that the formation of $Fe_2O_3$ tended to decrease with increasing calcination temperature.

Figure 4 shows the UV–vis absorbance spectra of the Fe-BLT powders calcined at different temperatures. UV-vis absorption spectroscopy was used to estimate the band gap of the samples. The optical band gap ($E_g$) of the samples was measured from the UV-vis absorbance spectra at wavelengths ranging from 200 to 800 nm. The optical band gap ($E_g$) was estimated using the method proposed by Wood and Tauc [21]. According to these authors, the optical band gap is associated with the absorbance and photon energy according to the following equation:

$$h\nu\alpha \propto (h\nu - E_g)^m,$$

where α is the absorbance, h is Planck's constant, ν is the frequency of the incident photon, $E_g$ is the optical band gap, and m is a constant associated with different types of electronic transitions. The $E_g$ value was obtained by extrapolating the linear segments of the curves towards the x-axis.

As a result, the optical band gap of iron-doped BLT samples increased gradually from 1.91 eV to 2.11 eV with increasing calcination temperature (300~900°C). In addition, decreases in the absorption intensity at a certain range (600 ~ 700 nm ≈ 1.8 ~ 2.0 eV) indicated a decrease in the concentration of $Bi_2O_3$, $Fe_2O_3$ or secondary $BiFeO_3$ phase in the XRD pattern. The optical band gaps of the samples heat treated at 300, 400, 500°C and 600°C were 1.91 eV, 1.93 eV, 1.95 eV and 1.97 eV, respectively, but there was incomplete formation on the orthorhombic structure in the amorphous and mixture state with $Bi_2O_3$ and $Fe_2O_3$ phases according to XRD. The optical band gap of the sample calcined at 700, 800 and 900 °C were increased slightly from 2.03 eV to 2.11 eV with a concomitant decrease in the secondary phase ($BiFeO_3$), as shown in Fig. 2. Although the samples calcined at 900°C showed the highest band gap among the three samples, only these samples formed without Fe-based secondary phases. The decrease in the optical band gap ($E_g$) was attributed to Fe substitution for Ti in BLT. Fe atoms are responsible for the modification of the electronic structure in the BLT-based oxides. These conditions can be correlated with the renormalization of the electronic band structure and band-gap. For example, J. S. Jang et al [22] reported that Fe substituted for Ti could play an important role in reducing the band gap in the $Zn_2TiO_4$ system under visible light irradiation. They reported that the band gap of $Zn_2TiO_4$ was reduced from 3.1 eV to 2.48 eV by Fe doping.

Consequently, the substitution of BLT by Fe by calcination at 700, 800°C and 900°C would be enough to control the band gap of BLT. The most appropriate calcination temperature was found to be 900°C without unnecessary absorption intensity for measurements of the band gap value. Table 1 presents the experimental results.

## IV. CONCLUSIONS

It was found that Fe atoms can decrease the optical band gap of BLT. On the other hand, XRD patterns of Fe-doped BLT calcined at 700℃ revealed secondary $BiFeO_3$ phases formed by Fe and Bi. To solve this problem, the Fe-BLT sample was prepared by calcination at temperatures greater than 700℃. As a result, the optimized calcination temperature of Fe-BLT was 900℃ due to the complete removal of $BiFeO_3$ phases due to a phase transition and decomposition at that temperature. Band gap control the thermal decomposition of a secondary phase is expected to be applicable to other complex oxide materials and provide a new tool for manipulating oxide optoelectronics for energy applications.

## ACKNOWLEDGMENTS

This study was supported by the Human Resources Development program (No. 20124030200010) under a Korea Institute of Energy Technology Evaluation and Planning (KETEP) grant funded by the Korea government's Ministry of Trade, Industry and Energy and by the Basic Science Research Program through the National Research Foundation of Korea (NRF) funded by the Ministry of Science, ICT & Future Planning (No. 2013005417).

## REFERENCES

[1] M. Qin, K. Yao, and Y. C. Liang, Appl. Phys. Lett. **93**, 122904 (2008).

[2] J. Xing, K. J. Jin, H. B. Lu, M. He, G. Z. Liu, J. Qiu, and G. Z. Yang, Appl. Phys. Lett. **92**, 071113 (2008).

[3] Y. S. Yang, S. J. Lee, S. Yi, B. G. Chae, S. H. Lee, H. J. Joo, and M. S. Jang, Appl. Phys.

Lett. **76**, 774 (2000).

[4] J. F. Scott, Science. **315**, 954 (2007).

[5] M. Alexe, and D. Hesse, Nat. Commun. **2**, 256 (2011).

[6] S. Y. Yang, L. W. Martin, S. J. Byrnes, T. E. Conry, S. R. Basu, D. Paran, L. Reichertz, J. Ihlefeld, C. Adamo, A. Melville, Y.-H. Chu, C.-H. Yang, J. L. Musfeldt, D. G. Schlom, J. W. Ager III and R. Ramesh, Appl. Phys. Lett. **95**, 062909 (2009).

[7] Y. Hou, J. Q. Xue, Z. M. Huang, T. X. Li, Y. J. Ge, J. H. Chu, Thin Solid Films. **517**, 901-904 (2008).

[8] Z. G. Hu, G. S. Wang, Z. M. Huang, J. H. Chu, J. Appl. Phys. **93**, 3811 (2003).

[9] S. H. Hu, J. Chen, Z. G. Hu, G. S. Wang, X. J. Meng, J. H. Chu, N. Dai, Mater. Res. Bull. **39**, 1223–1229 (2004).

[10] W. S. Choi, M. F. Chisholm, D. J. Singh, T. Choi, G. E. Jellison, and H. N. Lee, Nat. Comm. **3**, 689 (2012).

[11] N. Kumar, A. Kaushal, C. Bhwardwaj, D. Kaur, Optoelectron. Adv. Mater.-Rapid Commun. **4**, 10, 1497 - 1502 (2010).

[12] W. S. Choi and H. N. Lee, Appl. Phys. Lett. **100**, 132903 (2012).

[13] J. Y. Han, and C. W. Bark, to be published in Mol. Cryst. Liq. Cryst (2014).

[14] C. W. Bark, Met. Mater. Int. **19**, 1361-1364 (2013).

[15] M. Gotic, S. Popovic, and S. Music, Mater. Lett. **61**, 709-714 (2007).

[16] R. Nowosielski, G. Dercz, L. Pająk and R. Babilas, J. Achiev. Mater. Manuf. Eng. **17**, 117-120 (2006).

[17] M. F. R. Fouda, M. B. ElKholy, S. A. Mostafa, A. I. Hussien, M. A. Wahba, and M. F. El-Shahat, Adv. Mat. Lett. **4**, 5, 347-353 (2013).

[18] P. C Sati, M. Arora, S. Chauhan, S. Chhoker and M. Kumar, J. Phys. Chem. Solids. **75**, 105–108 (2014).

[19] D. C. Arnold, K. S. Knight, F. D. Morrison, and P. Lightfoot, Phys. Rev. Lett. **102**, 027602 (2009).

[20] D. C. Arnold, K. S. Knight, G. Catalan, S. A. T. Redfern, J. F. Scott, P. Lightfoot, and F. D. Morrison, Adv. Funct. Mater. **20**, 2116 (2010).

[21] D. L. Wood, J. Tauc, Phys. Rev. B. **5**, 3144 (1972).

[22] J. S. Jang, P. H. Borse, J. S. Lee, K. T. Lim, O. S. Jung, E. D. Jeong, J. S. Bae, M. S. Won and H. G. Kim, Bull. Korean Chem. Soc. **30**, 30212009 (2009).

Table. 1. Comprehensive experimental results from XRD and the $E_g$ values.

| Calcination Temperature(°C) | Result Phases | Band gap ($E_g$) |
|---|---|---|
| 300°C | | 1.91 eV |
| 400°C | α-$Bi_2O_3$ + α-$Fe_2O_3$ +$TiO_2$ phase | 1.93 eV |
| 500°C | | 1.95 eV |
| 600°C | α-$Bi_2O_3$ + α-$Fe_2O_3$ + Fe-BLT + $BiFeO_3$ phase | 1.97 eV |
| 700°C | Fe-BLT + $BiFeO_3$ phase | 2.03 eV |
| 800°C | | 2.1 eV |
| 900°C | Fe-BLT phase | 2.11 eV |

Figure Captions.

Fig. 1. Flow chart showing the procedure for synthesizing the samples at different calcination temperatures.

Fig. 2. XRD patterns of the $Bi_{3.25}La_{0.75}FeTi_2O_{12}$ (Fe-BLT) samples calcined at 300, 400, 500, 600, 700, 800, and 900°C. The vertical lines indicate the position of ICSD No. 150091.

Fig. 3. SEM images of the $Bi_{3.25}La_{0.75}FeTi_2O_{12}$ (Fe-BLT) powders calcined at (a) 300°C, (b) 400°C, (c) 500 °C, (d) 600 °C, (e) 700 °C, (f) 800°C, (g) 900°C.

Fig. 4. UV-vis absorbance spectra of the $Bi_{3.25}La_{0.75}FeTi_2O_{12}$ (Fe-BLT) powders calcined at (a) 300°C, (b) 400°C, (c) 500°C, (d) 600°C, (e) 700°C, (f) 800°C, (g) 900°C for 2.5 h in air. The extrapolation lines (dashed lines) indicate the optical band gap.

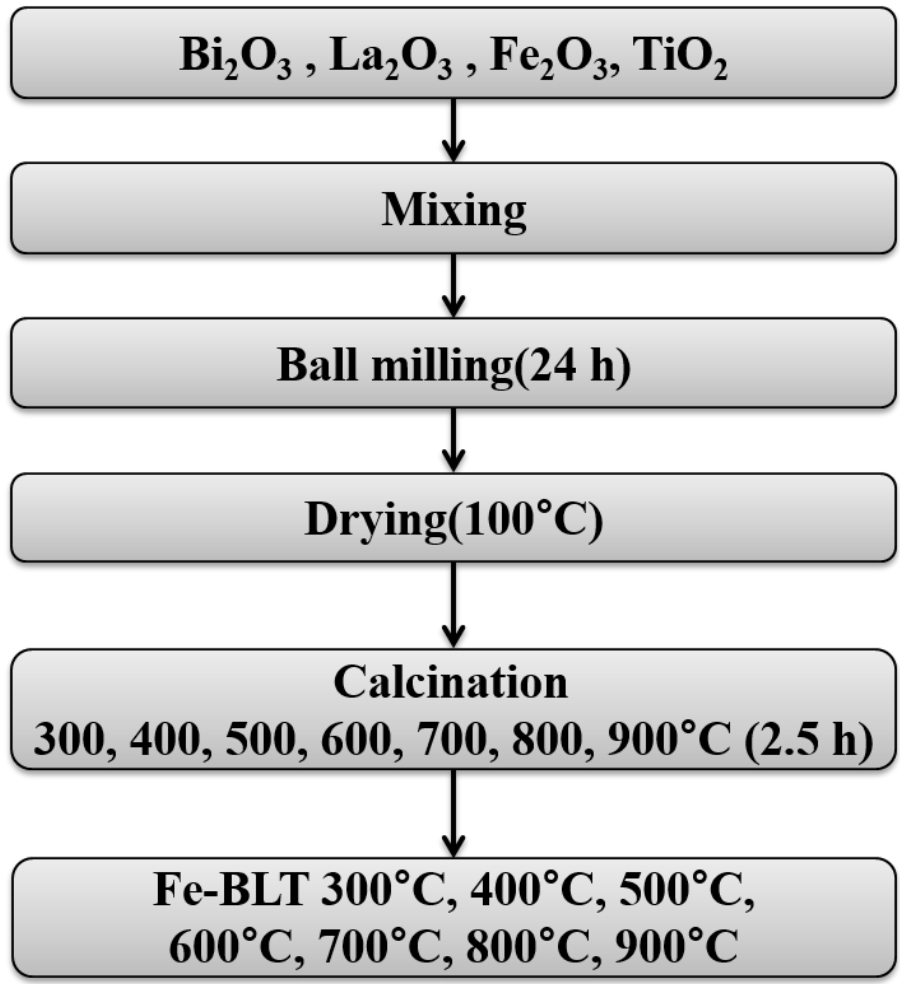

Figure 1.

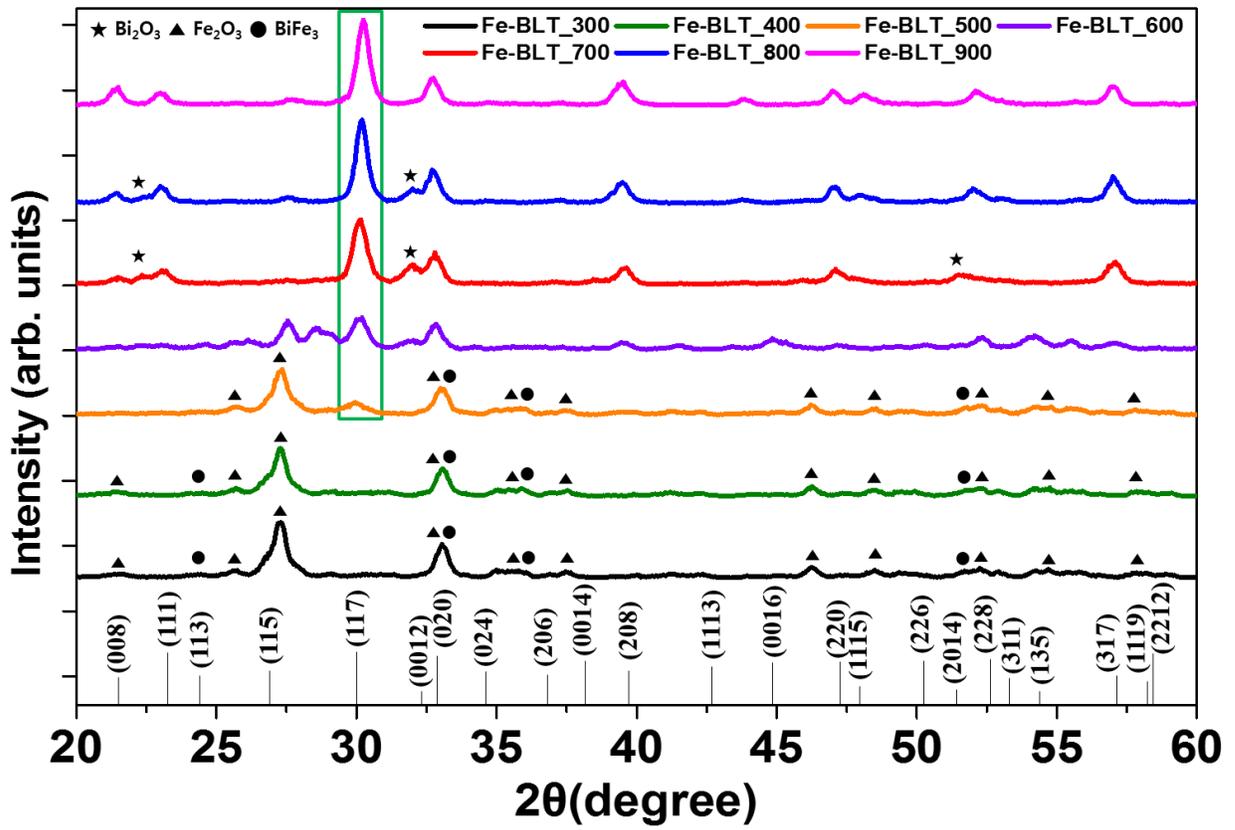

Figure 2.

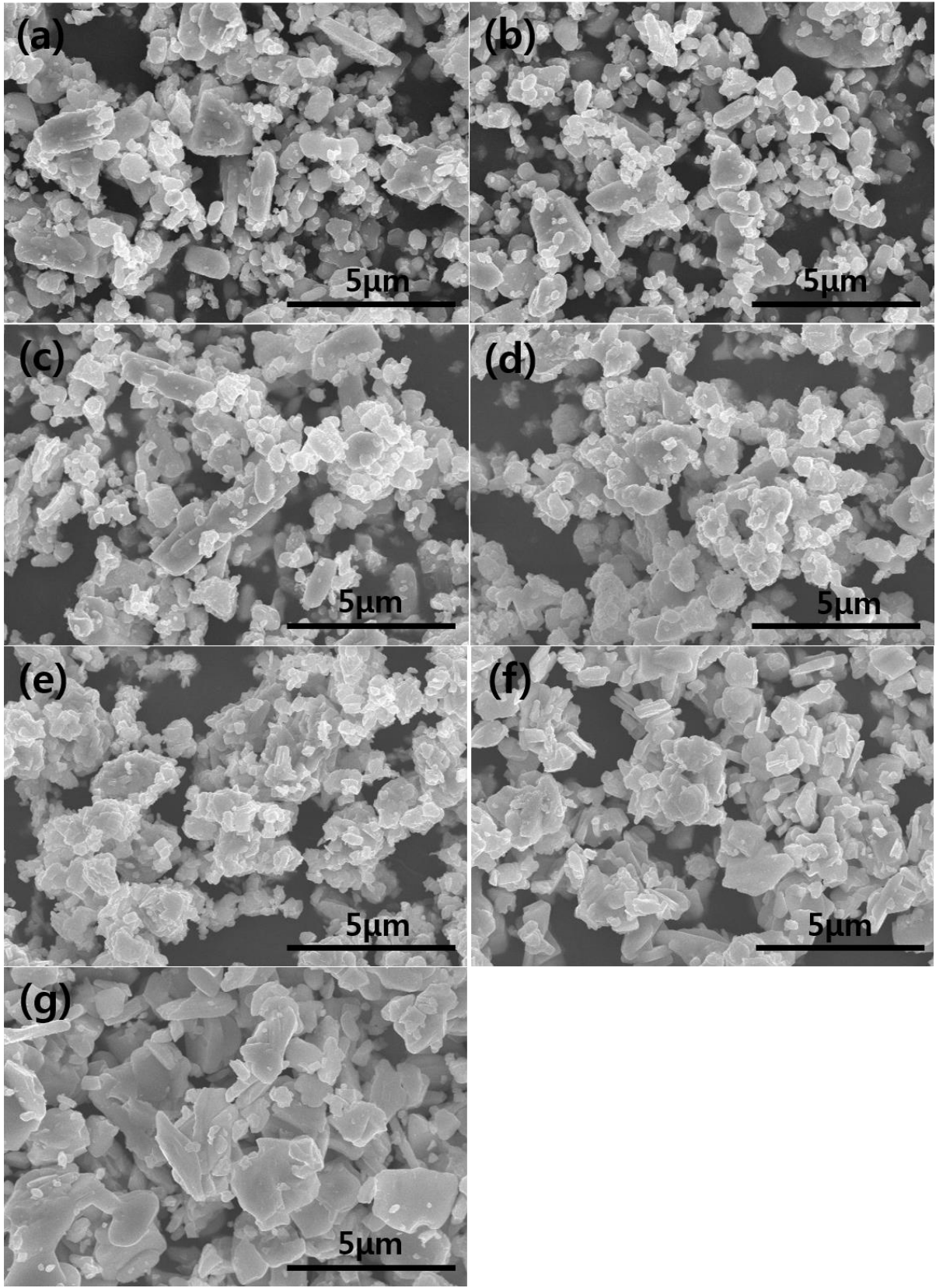

Figure 3.

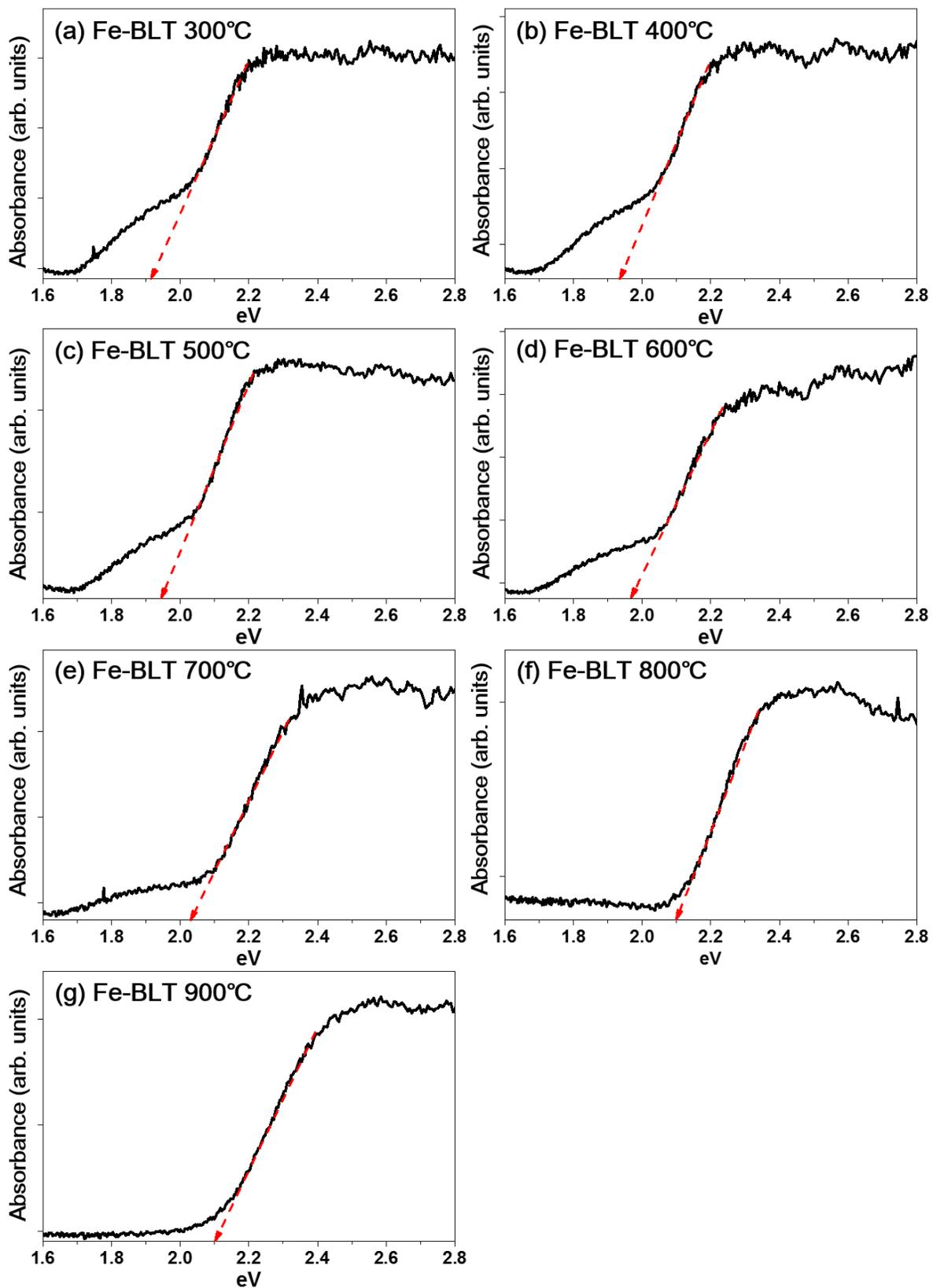

Figure 4.